\documentclass[aps,prl,twocolumn,superscriptaddress,apsrev]{revtex4-1}

\usepackage{graphicx}
\usepackage{epsfig}
\usepackage{amsmath,amsthm}
\usepackage{color}
\usepackage{verbatim}
\begin{document}

\title{Spin Mapping of Surface and Bulk Rashba States in Ferroelectric $\alpha$-GeTe(111) Films}
\author{H. J. Elmers}\email{elmers@uni-mainz.de}
\affiliation{Institut f\"{u}r Physik, Johannes Gutenberg-Universit\"{a}t, Staudingerweg 7, D-55099 Mainz, Germany}
\author{R. Wallauer}
\affiliation{Institut f\"{u}r Physik, Johannes Gutenberg-Universit\"{a}t, Staudingerweg 7, D-55099 Mainz, Germany}
\author{M. Liebmann}
\affiliation{II. Physikalisches Institut B and JARA-FIT, RWTH Aachen University, D-52074 Aachen, Germany}
\author{J. Kellner}
\affiliation{II. Physikalisches Institut B and JARA-FIT, RWTH Aachen University, D-52074 Aachen, Germany}
\author{M. Morgenstern}
\affiliation{II. Physikalisches Institut B and JARA-FIT, RWTH Aachen University, D-52074 Aachen, Germany}
\author{R.N. Wang} 
\affiliation{Paul-Drude-Institut f\"{u}r Festk\"{o}rperelektronik, Hausvogteiplatz 5-7, 10117 Berlin, Germany}
\author{J.E. Boschker}
\affiliation{Paul-Drude-Institut f\"{u}r Festk\"{o}rperelektronik, Hausvogteiplatz 5-7, 10117 Berlin, Germany}
\author{R. Calarco}
\affiliation{Paul-Drude-Institut f\"{u}r Festk\"{o}rperelektronik, Hausvogteiplatz 5-7, 10117 Berlin, Germany}
\author{O. Rader}
\affiliation{Helmholtz-Zentrum Berlin f{\"u}r Materialien und Energie, Elektronenspeicherring BESSY II, Albert-Einstein-Str. 15, D-12489 Berlin, Germany}
\author{D. Kutnyakhov}
\affiliation{Institut f\"{u}r Physik, Johannes Gutenberg-Universit\"{a}t, Staudingerweg 7, D-55099 Mainz, Germany}
\author{S.V. Chernov}
\affiliation{Institut f\"{u}r Physik, Johannes Gutenberg-Universit\"{a}t, Staudingerweg 7, D-55099 Mainz, Germany}
\author{K. Medjanik}
\affiliation{Institut f\"{u}r Physik, Johannes Gutenberg-Universit\"{a}t, Staudingerweg 7, D-55099 Mainz, Germany}
\author{C. Tusche}
\affiliation{Max Planck Institute for  Microstructure Physics, Weinberg 2, D-06120 Halle, Germany.}
\author{M. Ellguth}
\affiliation{Max Planck Institute for  Microstructure Physics, Weinberg 2, D-06120 Halle, Germany.}
\author{H. Volfova}
\affiliation{Department Chemie, Ludwig-Maximilians-Universit{\"a}t M{\"u}nchen, 81377 Munich, Germany}
\author{J. Braun}
\affiliation{Department Chemie, Ludwig-Maximilians-Universit{\"a}t M{\"u}nchen, 81377 Munich, Germany}
\author{J. Minar}
\affiliation{Department Chemie, Ludwig-Maximilians-Universit{\"a}t M{\"u}nchen, 81377 Munich, Germany}
\author{H. Ebert}
\affiliation{Department Chemie, Ludwig-Maximilians-Universit{\"a}t M{\"u}nchen, 81377 Munich, Germany}
\author{G. Sch{\"o}nhense}
\affiliation{Institut f\"{u}r Physik, Johannes Gutenberg-Universit\"{a}t, Staudingerweg 7, D-55099 Mainz, Germany}


\begin{abstract}
A comprehensive  mapping of the spin polarization of the electronic bands in ferroelectric $\alpha$-GeTe(111) films
 has been performed using a time-of-flight momentum microscope equipped with an imaging spin filter that enables a simultaneous 
measurement of more than 10.000 data points (voxels).
A Rashba type splitting of both surface and bulk
bands with opposite spin helicity of the inner and outer Rashba bands is found
revealing a complex spin texture at the Fermi energy.
The switchable inner electric field of GeTe implies new functionalities for spintronic devices.
\end{abstract}

\maketitle

The strong coupling of electron momentum and spin in low-dimensional structures
 allows an electrically
controlled spin manipulation in spintronic devices~\cite{Wolf2001,Zutic2004,Maekawa2006,Awschalom2007},
e.g. via the Rashba effect~\cite{Rashba1960}.

The Rashba effect has first been experimentally demonstrated in semiconductor heterostructures,
where an electrical field perpendicular to the layered structure, i.e. perpendicular
to the electron momentum, determines the electron spin orientation relative to its 
momentum~\cite{Datta1990,Nitta1997,Engels1997}.
An asymmetric interface structure causes the necessary  inversion symmetry breaking
and accounts for the special spin-splitting of electron states, the Rashba effect~\cite{Rashba1960},
the size of which can be tuned by the strength of the electrical field.

For most semiconducting materials the Rashba effect causes only a quite small splitting of the order of $10^{-2}$~\AA$^{-1}$
and thus requires experiments at very low temperatures~\cite{Koo2009,Wunderlich2010,Betthausen2012} and also implies
large lateral dimensions for potential spintronic applications. 
A considerably larger splitting  has been predicted theoretically~\cite{DiSante2013}
and was recently found experimentally for the surface states of GeTe(111)~\cite{Liebmann2015,Krempasky2015}.

GeTe is a ferroelectric semiconductor with a Curie temperature of 700~K.
Thus, besides the interface induced Rashba splitting,
the ferroelectric properties also imply a broken inversion symmetry within the bulk
and thus would allow for the electrical tuning of the bulk Rashba splitting
via switching the ferroelectric polarization~\cite{DiSante2013,Dresselhaus1955,Zhang2014}.
This effect is of great interest for non-volatile spin orbitronics~\cite{Wunderlich2010}.

For GeTe a bulk Rashba splitting of $0.19$~\AA$^{-1}$ 
has been predicted theoretically~\cite{DiSante2013}.
Experimentally, bulk-Rashba bands are rare and have only been found in the layered polar semiconductors
BiTeCl and BiTeI~\cite{Ishizaka2011,Lee2011,Wang2013,Sakano2013} that, however, are not switchable.

A characterization of the ferroelectric properties and a measurement of the spin polarization
of the surface states of GeTe(111) at selected $k$-points has been performed previously by 
force microscopy~\cite{Kolobov2014,Wang2014} and spin-resolved angular resolved photoemission spectroscopy, respectively~\cite{Liebmann2015}.
A recent experimental and theoretical study revealed that at the Fermi level the hybridization of surface and bulk states
causes surface-bulk resonant states resulting in unconventional spin topologies with chiral symmetry~\cite{Krempasky2015}.

Here, we demonstrate the spin structure of surface and bulk bands of the GeTe(111)
surface using the novel photoemission technique of spin-resolved time-of-flight momentum microscopy
that allows for an effective parallel spin mapping of the electronic states~\cite{Schonhense2015}.
The previous findings for the spin texture at the Fermi level are confirmed.
In addition we reveal a Rashba-type spin texture of the bulk band, too.

Single-domain epitaxial $\alpha$-GeTe(111) films 
with a thickness of 32 nm are grown by molecular beam epitaxy (MBE) on in-situ prepared Si(111) 
and are transferred in ultra-high vacuum (UHV) to the spin-resolved time-of-flight momentum microscope.
Thus, we are able to probe the pristine surface, which shows an
outward ferroelectric polarization. It was previously shown that a metastable inverted polarization
state can be written using ex-situ piezo force microscopy.
Details of the structural and ferroelectric properties of the samples are reported in Ref.~\cite{Liebmann2015}.

The photoemission experiment has been performed at the undulator beamline U125-NIM
at BESSY II in the single bunch mode, providing vacuum ultraviolet photon pulses (photon energy range 15~eV to 27~eV)
of 50~ps duration and a repetition rate of 1.25~GHz.
The time-of-flight momentum microscope yields a simultaneous acquisition of the 
$E(k_{||})$ spectral function in the full surface Brillouin zone and several eV energy interval.
Details of this instrument are given in Ref.~\cite{Schonhense2015}.
For the present experiment the overall energy and the $k_{||}$ resolution 
is 86~meV  and 0.03~\AA$^{-1}$.
The photon beam shines at an incidence angle of 68$^0$ onto the surface
parallel to the $\overline{\Gamma}$-$\overline{M}$ direction with linear polarization
with the electric field vector parallel to the plane of incidence (p-polarization).
Spin detection was achieved by electron reflection at the W(100) spin filter crystal
in the [010] azimuth at 45$^0$ reflection angle~\cite{Kolbe2011,Kutnyakhov2015}.  
The spin direction parallel to the $\overline{\Gamma}$-$\overline{M}$ direction is probed.
Two datasets $I^{\sigma}(E,k_{||})$, $\sigma=+,0$, were acquired at scattering energies 26.5~eV and 30.5~eV, respectively.
At 26.5~eV the reflection asymmetry is $A=0.3$~\cite{Kutnyakhov2015}  while it is negligibly small at 30.5~eV.
The spin polarization is then determined by 
\begin{equation}
P(E,k_{||})=\frac{2}{A} \frac{I^{+}(E,k_{||}) - r(E)I^{0}(E,k_{||})}{I^{+}(E,k_{||}) + r(E)I^{0}(E,k_{||})},
 \end{equation}
where $r(E)$ denotes the energy-dependent relative reflection coefficient as described in Ref.~\cite{Tusche2013}.
The energy dependence of $A$ was neglected as we analyzed only a narrow energy interval of less than 1~eV.

\begin{figure}
\includegraphics[width=\columnwidth]{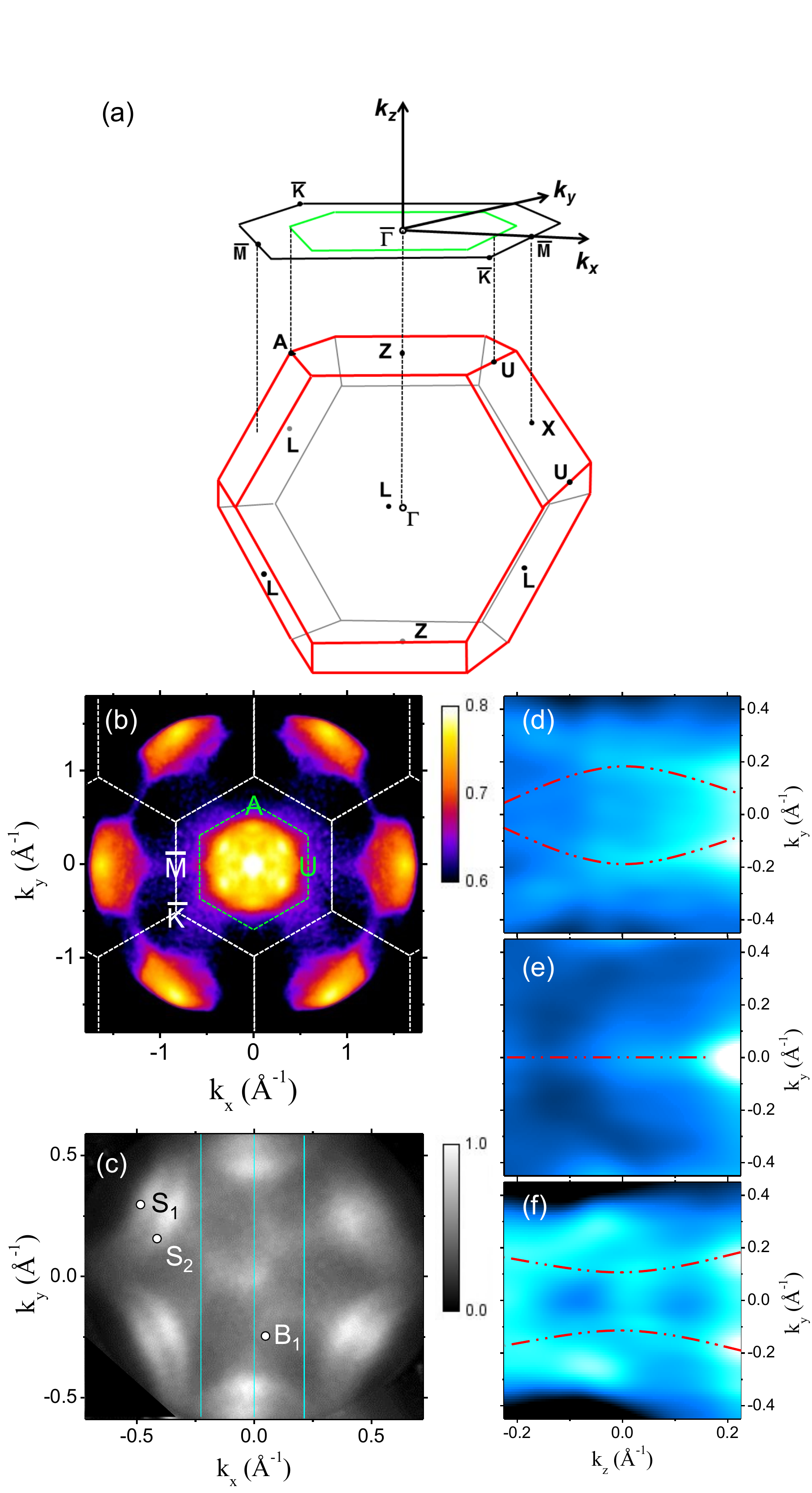}
\caption{\label{fig:Figure1} (Color online)
(a) Bulk and surface Brillouin zone of GeTe with marked high symmetry points and  
the projection of the hexagonal face of the Brillouin zone encircled in green. 
(b) Constant energy slice $I(E_F,k_x,k_y)$ of the photoelectron intensity distribution
for a photon energy of $h\nu=18.5$~eV covering the complete bulk Brillouin zone projection.
(c) Constant energy slice $I(E_b,k_x,k_y)$ with $E_b=0.4$~eV for $h\nu=22$~eV
with larger magnification. The two surface bands (S$_{1,2}$) and the bulk band B$_1$ are marked. 
Lines at konstant $k_x$ indicate position of cuts shown in (d-f).
(d-f)  Photoelectron intensity $I(0.4 {\rm eV}, k_x, k_y, k_z)$ for fixed values $k_x=0$ and $k_x=\pm 0.2$~\AA$^{-1}$.
$k_z$ is measured relative to the Z point in $k$-space, which is probed for $h\nu =19.5$~eV.
}
\end{figure}

First, we describe experimental results obtained  with the straight branch of the instrument without spin detection.
Fig.~\ref{fig:Figure1}(a) shows a sketch of the surface and bulk first Brillouin zone of GeTe(111).
The known six-fold symmetric shape of the bulk Rashba band~\cite{DiSante2013} is displayed on the ZAU 
hexagonal surface of the bulk Brillouin zone. 
Fig.~\ref{fig:Figure1} (b) presents a constant energy slice at the Fermi level $E=E_F$ covering the complete
first bulk Brillouin zone measured with a photon energy of 18.5~eV.
According to Ref.~\cite{Liebmann2015}
$h\nu=18.5$~eV probes the bulk bands at $E_F$ in the vicinity of the 
ZAU plane. 

Fig.~\ref{fig:Figure1}(c) shows a magnified constant energy spectrum at a binding energy
$E_b=0.4$~eV measured at a photon energy of $h\nu=22$~eV.
The splitted surface Rashba bands with a six-fold symmetry a clearly resolved. 
Additional arcs appear revealing a three-fold symmetry. In order to proof that
these arcs are bulk bands we performed the experiment for a number of photon energies
from $h\nu=15.5$~eV to $h\nu=24$~eV in steps of 0.5~eV resulting
in the four-dimensional dataset $I(E_b, k_x, k_y, k_z)$.
Each of these sets consisting of 1250 resolved $k$-points and 24 resolved energy-slices
took about 15 minutes of acquisition time.
From the data we extracted constant energy maps $I(0.4 {\rm eV},k_x,k_y,k_z)$ at a binding energy $E_b=0.4$~eV
 with fixed values $k_x=0$ and $k_x=\pm 0.2$~\AA$^{-1}$
[Figs.~\ref{fig:Figure1}(d-f)].
Bulk bands are identified by their dispersion with $k_z$ for Figs.~\ref{fig:Figure1}(d,f).
The symmetric behavior of the dispersion allows the conclusion
that the high symmetry point corresponds to a photon energy of 19.5~eV.
Using the known radius
of the Brillouin zone $\Gamma Z=0.91$~\AA$^{-1}$ and assuming an 
excitation into free electron states, we derive an inner potential of $E_F-E_i=9.2$~eV, 
agreeing well with  theoretical values~\cite{Liebmann2015}.
In contrast, the dispersion-less behavior of the band shown in Fig.~\ref{fig:Figure1}(e) 
indicates its nature as a surface or surface-resonant band.

\begin{figure}
\includegraphics[width=\columnwidth]{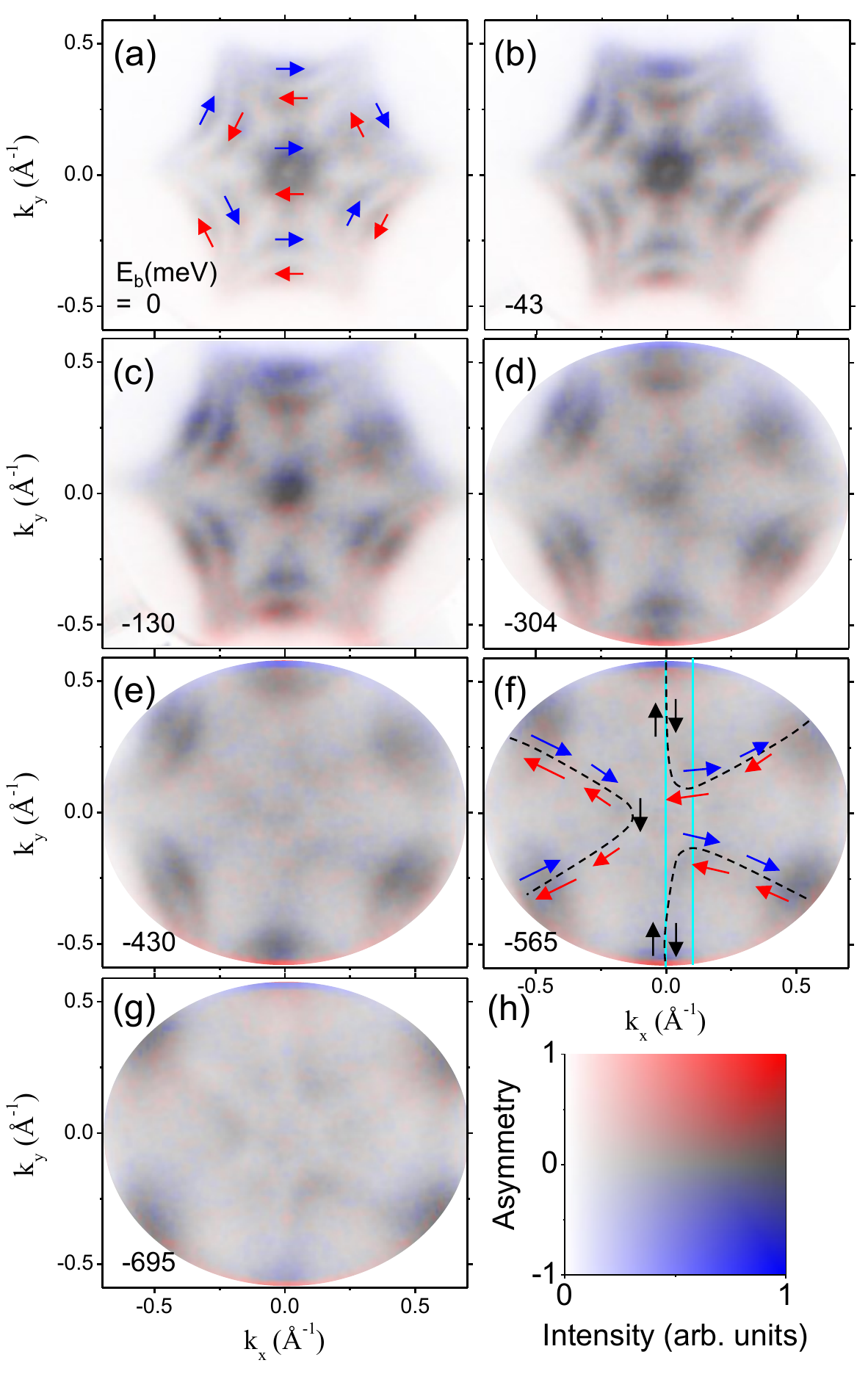}
\caption{\label{fig:Figure2} (Color online)
(a-g) Constant energy slices $P_x(E_b,k_x,k_y)$ of the photoelectron spin polarization component
along the $x$-axis combined with the corresponding intensity distribution
measured without spin resolution for $h\nu=22$~eV.
Each slice averages over an energy interval of 43~meV.
The corresponding color code for $P_x$  and $I$ is indicated in (h).
Arrows in (a) and (f) mark the deduced spin direction assuming a negligible perpendicular spin component.
}
\end{figure} 

\begin{figure}
\includegraphics[width=0.7\columnwidth]{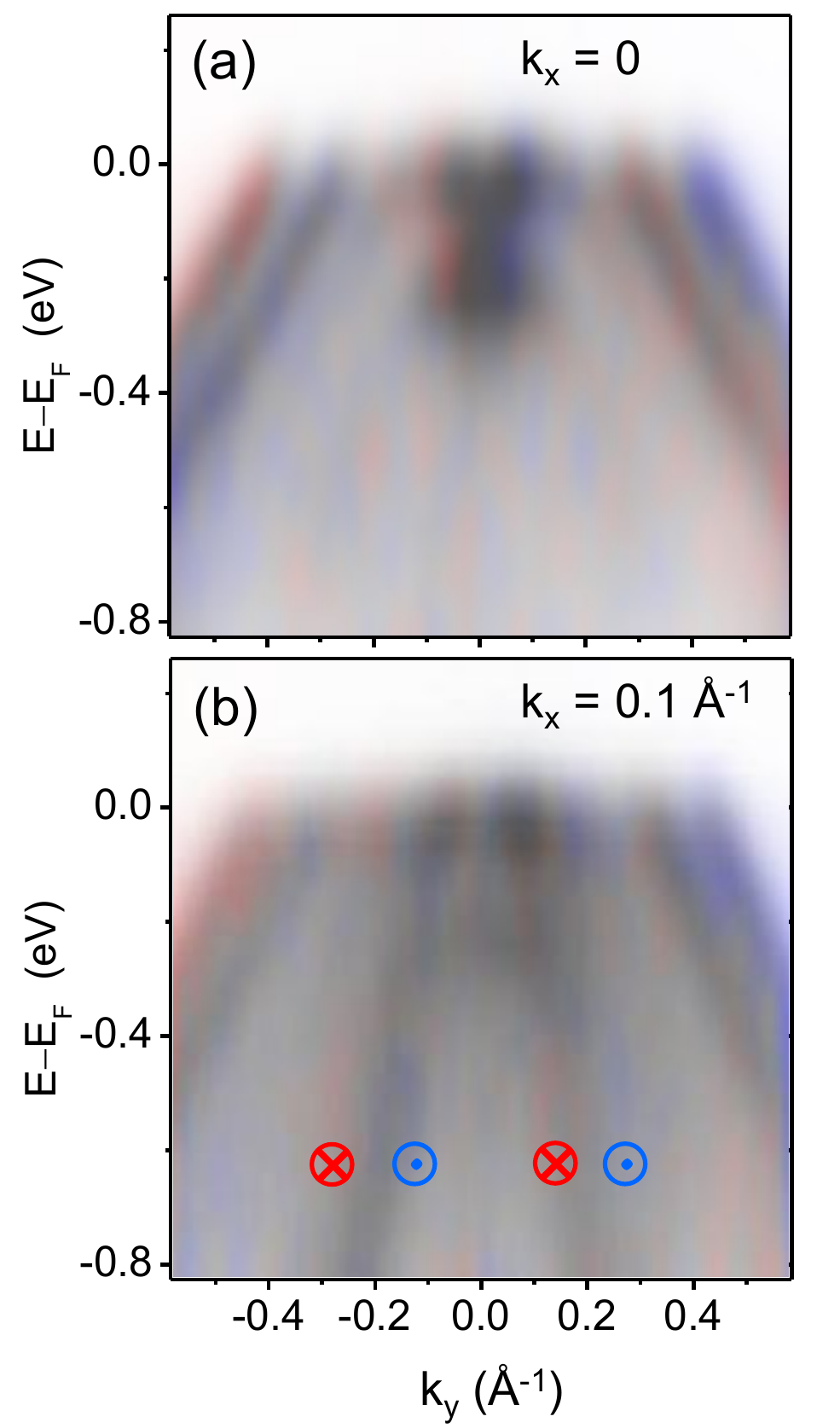}
\caption{\label{fig:Figure3} (Color online)
(a-b) Constant momentum cut $I(E_b,k_x,k_y)$ for $k_x=0$ and 
$k_x=+0.09$~\AA$^{-1}$ from the same dataset as shown in Fig.~\ref{fig:Figure2}.
The cuts are averaged over a momentum interval of $\Delta k_x=0.06$~\AA$^{-1}$.
}
\end{figure}

In the following, we discuss the spin polarization $P$ of the electronic bands measured
after reflection at the W(100) spin filter crystal probing the $x$-component $P_x$.
In order to remove artificial experimental asymmetries
we exploit the fact that the ($\Gamma$, Z, M)-plane represents
a mirror plane of the whole experiment enforcing the condition $P_x(E_b,k_x,k_y)=-P_x(E_b,k_x,-k_y)$.
From the complete three-dimensional $P_x(E_b,k_x,k_y)$ map we present constant energy slices
at selected binding energies in Fig.~\ref{fig:Figure2}(a-g). 
Each slice integrates over an energy interval of 43~meV.
At $E_F$ (Fig.~\ref{fig:Figure2}(a) ) the spin orientation is 
oriented clockwise (counter-clockwise) for the outer (inner) surface Rashba band
as seen from above the sample surface. 
The maximum value $P_x=0.7\pm0.1$ is observed
at the top and bottom of the outer Rashba surface band.   
The left and right outer Rashba surface bands show a polarization of $P_x=0.35\pm0.1$
in agreement with the Rashba-type spin texture, i.e. $\vec{P} \perp \vec{k}_{||}$.
Spin polarization for the inner bands are of opposite direction and only slightly smaller.
These results confirm previously published data~\cite{Liebmann2015,Krempasky2015}.

The constant energy slice at $E-E_F=-0.565$~eV shown in Fig.~\ref{fig:Figure2}(f)
also reveals the spin orientation of the bulk band although the Rashba splitting
is not resolved in the intensity signal.
In order to emphasize the weak spin contrast we present the $E(k_y)$ cuts in Fig.~\ref{fig:Figure3}(a,b)
parallel to ZA  for $k_x=0$ and $k_x=0.09$~\AA$^{-1}$, the latter representing
a cut through the bulk band. 
The spin polarization is of opposite sign to the left and right from the 
bulk band's maximum intensity.
The maximum spin polarization $P_x=0.12\pm0.1$ of the bulk band is considerably smaller   than that of the surface bands.
This can be explained by the smaller Rashba splitting and the corresponding partial overlap
of the intensity of inner and outer band.
A large perpendicular spin component cannot be excluded from the experiment but seems
unlikely in respect to the theoretical predictions~\cite{Liebmann2015,Krempasky2015}.
The spin direction is mostly perpendicular to the $E(k_y)$-cut shown in Figs.~\ref{fig:Figure3}(a,b).
The $P_x$ component for the larger absolute $k_{||}$ of the Rashba splitted band is also parallel to 
that of the surface band with larger    $k_{||}$ value, which is in agreement with the 
Rashba-type spin orientation, respective the theoretical predictions for the equilibrium ferroelectric polarization
of GeTe~\cite{Dresselhaus1955,Krempasky2015}.

In summary, we performed a spin mapping of surface and bulk Rashba states in the complete surface and bulk Brillouin zone of
ferroelectric $\alpha$-GeTe(111) films.
The theoretical prediction of the Rashba splitting of bulk bands is confirmed.
The spin helicity is in line with the prediction for the equilibrium
outwards electrical polarization of GeTe(111)
connecting it to the ferroelectricity of GeTe.
This  represents the necessary precondition 
for application of this material in non-volatile spintronic devices.

The authors gratefully acknowledge financial support from the 
BMBF (05K13UM1, 05K13PA and 05K13WMA),
Deutsche 
Forschungsgemeinschaft (DFG EL/18-1 within SPP1666 and SFB 917-A3),
Center of INnovative Emerging MAterials (CINEMA) and Stiftung Rheinland
Pfalz f{\"u}r Innovation (project 1038).

\bibliography{GeTe111}

\end{document}